\documentclass[aps,prc,twocolumn,superscriptaddress,showpacs,showkeys]{revtex4}
\usepackage{amsmath,amssymb}
\usepackage{bm}
\usepackage[dvips]{graphicx}
\begin{document}

\title{$\Lambda(1405)$ production in the $\pi^-p\to K^0\pi\Sigma$
reaction}

\author{T.~Hyodo}%
\affiliation{Research Center for Nuclear Physics (RCNP), 
Ibaraki, Osaka 567-0047, Japan}%

\author{A.~Hosaka}%
\affiliation{Research Center for Nuclear Physics (RCNP), 
Ibaraki, Osaka 567-0047, Japan}%

\author{E.~Oset}%
\affiliation{Departmento de F\'isica Te\'orica and IFIC,
Centro Mixto Universidad de Valencia-CSIC,
Institutos de Investigaci\'on de Paterna, Aptd. 22085, 46071
Valencia, Spain}%

\author{A.~Ramos}%
\affiliation{Department d'Estructura i Constituents de la Mat\`eria,
Universitat de Barcelona, Diagonal 647, 08028 Barcelona, Spain}%

\author{M.~J.~Vicente~Vacas}%
\affiliation{Departmento de F\'isica Te\'orica and IFIC,
Centro Mixto Universidad de Valencia-CSIC,
Institutos de Investigaci\'on de Paterna, Aptd. 22085, 46071
Valencia, Spain}%

\date{\today}

\begin{abstract}
    We discuss the mechanisms that lead to $\Lambda(1405)$ production
    in the $\pi^-p\to K^0\pi\Sigma$ reaction.
    The problem has gained renewed interest after different works
    converge to the conclusion that there are two
    resonances around the region of 1400 MeV,
    rather than one, and that they couple
    differently to the $\pi\Sigma$ and $\bar{K}N$ channels.
    We look at the dynamics of that reaction and find two mechanisms
    which eventually filter each one of the resonances,
    leading to very different shapes of the $\pi\Sigma$ invariant
    mass distributions.
    The combination of the two mechanisms leads to a shape of this
    distribution compatible with the experimental measurements.
\end{abstract}


\maketitle

\section{Introduction}

The $\Lambda(1405)$ resonance has kept the attention of researchers
for a long time.
It has been suggested to be a quasi-bound $\bar{K}N$
state which appears in a coupled channel approach to the strangeness
$S=-1$ meson-baryon interaction close to the $\bar{K}N$
threshold~\cite{Jones:1977yk}.
Subsequent studies combining chiral dynamics and unitarity in coupled
channels~\cite{Kaiser:1995eg,Kaiser:1997js,Oset:1998it}, have
brought new light into this idea and
substantiated it from the modern perspective of chiral Lagrangians.
Yet, one of the theoretical surprises along these lines has been
the persistent finding of two
resonances close to the nominal $\Lambda(1405)$.
It was already found in Ref.~\cite{Fink:1990uk} using the cloudy bag 
model that two poles, rather than one, appeared in the region around 
1405 MeV with the quantum numbers of $\Lambda(1405)$,
$I(J^P)=0(1/2^-)$.
Subsequently, two similar poles have been found in
Refs.~\cite{Oller:2000fj,Oset:2001cn,Jido:2002yz,Garcia-Recio:2002td,
Hyodo:2002pk,Hyodo:2003qa,Garcia-Recio:2003ks,Nam:2003ch} using
chiral unitary approaches to the problem. All the different
approaches agree qualitatively in the following :
\begin{itemize}
    \item[a)]  Two poles appear in the complex plane close to the 1405
    MeV region.

    \item[b)]  The pole at lower energies has a larger width than the one
    appearing at higher energies.

    \item[c)]  The lower energy resonance couples strongly to $\pi\Sigma$
    and weakly to $\bar{K}N$, while the opposite occurs for the
    resonance appearing at higher energy,
    although in models which do not fit the $K^-p$ threshold
    branching ratio~\cite{Garcia-Recio:2002td},
    the couplings look more similar.
\end{itemize}

A clarification of this interesting result has been made in
Ref.~\cite{Jido:2003cb} where the two $\Lambda(1405)$ states have
been interpreted in the following way :
The SU(3) decomposition of the octet representation of the $1/2^+$
baryons times the octet of the $0^-$ pseudoscalar mesons leads to a
singlet and two octets (apart from the $10$, $\bar{10}$ and $27$
representations).
The two octets appear degenerate in the limit of exact SU(3) symmetry,
but the explicit breaking of SU(3) due
to different masses of the mesons and the baryons breaks the
degeneracy.
When this happens, one observes two trajectories for the poles in
terms of a given SU(3) breaking parameter.
Two branches for $I=1$ and two for $I=0$ emerge, one of them moving
to the higher energy side of the SU(3) symmetric pole and the
other one moving to the lower energy side.
The $I=0$ branch that moves to low energies comes very close to the
singlet pole, in such a way that reactions occurring in the energy
region around 1400 MeV will excite both resonances, but only one
apparent bump will be seen, giving the impression that there is only
one resonance.
Yet, it was discussed in Ref.~\cite{Jido:2003cb} that, given the fact
that the two resonances couple very differently to the $\bar{K}N$
and $\pi\Sigma$ states, different reactions can give
more weight to one or the other resonance leading to different shapes
in the $\pi\Sigma$ mass distribution.
Examples were given there for possible situations and the $K^-p \to
\Lambda(1405)\gamma$ reaction was suggested as a means to see a case 
in which much weight is given to the higher mass resonance, 
resulting in a $\pi\Sigma$
mass distribution narrower than the nominal one with the peak position
shifted by about 20 MeV to higher energies.

The lesson learned in that paper is that the shape of the $\pi\Sigma$
mass distribution obtained for a certain reaction depends drastically
on the dynamics of the reaction.
This reopens a problem since the shape of the $\Lambda(1405)$ resonance
from the $\pi\Sigma$ mass distribution was formerly assumed to be an
intrinsic property of the resonance and hence independent of the
reaction used to produce it.
For instance, in Refs.~\cite{Kaiser:1995eg,Kaiser:1997js,
Oset:1998it,Garcia-Recio:2002td,Hyodo:2002pk,
Hyodo:2003qa,Nam:2003ch},
the $\pi\Sigma$ mass distribution was generated assuming
\begin{equation}
    \frac{d\sigma}{d M_I}=C|t_{\pi\Sigma\to\pi\Sigma}|^2p_{CM}
    \label{eq:mdist1}
\end{equation}
with $p_{CM}$ the momentum of the pion in the $\pi\Sigma$ rest frame.
In practice, this will not necessarily happen, and at least it will
not in some reactions. Indeed, if one bears in mind that the
$\Lambda(1405)$ resonance is built up from the multiple scattering of 
the coupled channels, $\bar{K}N$, $\pi\Sigma$, $\eta\Lambda$, $K\Xi$,
one can produce the resonance first by producing any of these
channels and then having final state interaction leading to the final
$\pi\Sigma$ state. Hence, instead of Eq.~\eqref{eq:mdist1}, we
should rather have
\begin{equation}
    \frac{d\sigma}{d M_I}=|\sum_{i}C_it_{i\to\pi\Sigma}|^2p_{CM}
    \label{eq:mdist2}
\end{equation}
with $i$ standing for any of the coupled channels, and the
coefficients $C_i$ will depend upon the particular reaction.
If there is one pole around $\Lambda(1405)$, then the shape of 
$|t_{i\to \pi\Sigma}|$ is almost uniquely determined independent of
the channel $i$.
However, when there are two poles, it depends on $i$,
since the different channel $i$ couples to the poles with different
strengths~\cite{Jido:2003cb}.
Therefore, the mass distribution develops one or another shape
depending on the coefficients $C_i$.
The fact that this distribution follows
Eq.~\eqref{eq:mdist2} rather than Eq.~\eqref{eq:mdist1}, was already 
pointed out in Ref.~\cite{Oller:2000fj}.
However, no attempt
was done to calculate the $C_i$ coefficients but rather
they were fitted to 
the data to obtain the experimental shape of the $\Lambda(1405)$
resonance.

The aim of the present work is to study the $\pi^-p\to K^0\pi\Sigma$ 
reaction, from which the experimental data of the $\Lambda(1405)$
resonance are usually extracted~\cite{Thomas:1973uh}.
Another source of experimental information comes from the
$K^-p\to\Sigma^{+}(1660)\pi^-$ reaction followed by
$\Sigma^+(1660)\to \Lambda(1405)\pi^+$, $\Lambda(1405)\to
\pi\Sigma$~\cite{Hemingway:1985pz}.
The purpose of the present 
paper is to investigate the dynamics that goes
into the $\pi^-p\to K^0\pi\Sigma$ reaction,
and obtain the
coefficients entering Eq.~\eqref{eq:mdist2} which determine the shape 
of the $\Lambda(1405)$.

In the next section, we explore the analogy of the $\pi^-p\to K^0\pi\Sigma$
reaction with the $\pi N\to \pi\pi N$ reaction making an SU(3)
extrapolation of the same low energy $\pi N\to \pi\pi N$ model.
In section 3, we show the results with this model,
and in section 4,
we  look at the contribution of resonance states
excited in the s-channel.
Results and conclusions are shown in
subsequent sections.

\section{Chiral amplitudes for the $\pi^-p\to K^0\pi\Sigma$ reaction}

\subsection{Lagrangian}

In this subsection, we briefly summarize the chiral Lagrangian,
that we use in the following calculations.
The meson-meson Lagrangian at the lowest order needed here takes on
the form~\cite{Gasser:1985gg,Meissner:1993ah}
\begin{equation}
{\cal L}_2 = \frac{1}{12 f^2} < ( \partial_\mu
\Phi \Phi - \Phi \partial_\mu \Phi)^2 + M \Phi^4 > 
\label{eq:MLag} \ ,
\end{equation}
where $f$ is the meson decay constant,
$M$ is the quark mass matrix $M=\text{diag}(\hat{m},\hat{m},m_s)$,
and the symbol $< \, >$ denotes trace of SU(3) matrices.
Similarly, following Refs.~\cite{Pich:1995bw,Ecker:1995gg,Bernard:1995dp}, 
we write the lowest order chiral Lagrangian including the
coupling of the octet of pseudoscalar mesons
to the octet of $1/2^+$ baryons as
\begin{equation}
    \begin{split}
	\mathcal{L}_1^{(B)} =&
	< \bar{B} i \gamma^{\mu} \nabla_{\mu} B> - M_B <\bar{B} B> \\
	&+\frac{1}{2} D <\bar{B} \gamma^{\mu} \gamma_5 
	\left\{ u_{\mu}, B \right\} > \\
	&+ \frac{1}{2} F <\bar{B} \gamma^{\mu} \gamma_5 [u_{\mu}, B]> \ ,
    \end{split}
    \label{eq:BLag}
\end{equation}
where we have adopted the usual definition for the baryon field $B$, 
mesonic current $u_\mu$ and the covariant derivative 
$\nabla_\mu$~\cite{Pich:1995bw}.  
The strengths of the $F$ and $D$ coupling constants are fixed as 
$F = 0.51,\; D=0.75$.  
At lowest order in momentum, that we will keep in our study, 
the meson-baryon interaction
Lagrangian comes from the $\Gamma_{\mu}$ term in the covariant derivative
of Eq.~\eqref{eq:BLag} :
\begin{equation}
    \mathcal{L}_{WT} = < \bar{B} i \gamma^{\mu} \frac{1}{4 f^2}
    [(\Phi \overleftrightarrow{\partial_{\mu}}\Phi) B
    - B (\Phi \overleftrightarrow{\partial_{\mu}}\Phi)] >
\label{eq:WTint} \ ,
\end{equation}
which gives the Weinberg-Tomozawa interaction.
Here $\Phi$ represents the octet meson field~\cite{Pich:1995bw}.  
The $D$ and $F$ terms of Eq.~\eqref{eq:BLag} provide the Lagrangian
where odd number of
mesons couple to the two baryons.
We derive the meson-baryon Yukawa interaction from this term as
\begin{equation}
    \begin{split}
	\mathcal{L}_{\text{Yukawa}}
	=&-\frac{1}{\sqrt{2}f} <
	D(\bar{B}\gamma^{\mu}\gamma_{5}
	\{\partial_{\mu}\Phi,B\}) \\
	&+F(\bar{B}\gamma^{\mu}\gamma_{5}
	[\partial_{\mu}\Phi,B]) >  \ ,
    \end{split}
    \label{eq:Yukawa}
\end{equation}
and the $MMMBB$ (three meson-two baryon) contact interaction  as
\begin{equation}
    \begin{split}
	\mathcal{L}_{\text{Contact}}
	=&\frac{1}{12\sqrt{2}f^3}<
	D(\bar{B}\gamma^{\mu}\gamma_{5}
	\{
	(\partial_{\mu}\Phi(\Phi^2) \\
	&-2\Phi\partial_{\mu}\Phi(\Phi)
	+\Phi^2\partial_{\mu}\Phi
	),B\}) \\
	&+F(\bar{B}\gamma^{\mu}\gamma_{5}
	[
	(\partial_{\mu}\Phi(\Phi^2) \\
	& -2\Phi\partial_{\mu}\Phi(\Phi)
	+\Phi^2\partial_{\mu}\Phi),B])> \ .
    \end{split}
    \label{eq:Contact}
\end{equation}

\subsection{Construction of the chiral amplitude}

We  construct a model for the $\pi^-p\to K^0\pi\Sigma$ reaction
at energies close to threshold for the $\pi^-p\to K^0\Lambda(1405)$
production.
This means a total center of mass energy $\sqrt{s}\sim 2$ GeV,
or equivalently a three momentum of the initial pion 
$p_{\pi}\sim 1.7\text{ GeV}/c$ in the Laboratory frame.

\begin{figure}[tbp]
    \centering
    \includegraphics[width=8cm,clip]{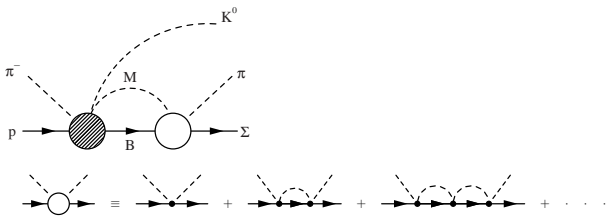}
    \caption{\label{fig:1}
    Diagrams entering the production of the $\Lambda(1405)$ in the
    $\pi^- p\to K^0\Lambda(1405) \to K^0\pi\Sigma$.
    In the figure, $M$ and $B$
    stand for the meson and baryon of the 10 possible
    coupled channels.
    The $\Lambda(1405)$ resonance 
    is dynamically generated by the final state
    interaction of $M$ and $B$.}
\end{figure}

Formally, we can separate the process into two parts.
The first one which involves tree level $\pi^-p\to K^0MB$ amplitudes,
and a second part which involves the final state
interaction $MB\to \pi\Sigma$,
which eventually generates a resonance if
kinematical and dynamical conditions allow for it.
This is shown in Fig.~\ref{fig:1}.
We produce dynamically the $\Lambda(1405)$ \textit{via}
the final state interaction, as it was done for the
photoproduction of the $\Lambda(1405)$ in the
$\gamma p\to K^{+}\Lambda(1405)$~\cite{Nacher:1998mi}.
This is accomplished by summing the series of diagrams depicted in
Fig.~\ref{fig:1} \textit{via} the Bethe-Salpeter equation in coupled
channels~\cite{Oset:1998it} 
\begin{equation}
    t=V+VGT
    \label{eq:BSE} \ ,
\end{equation}
with the kernel $V$ obtained from the lowest order chiral Lagrangians of
Eq.~\eqref{eq:WTint}.
The coupled channels appearing in this problem are 
$K^-p$, $\bar{K}^0n$, $\pi^0\Lambda$, $\pi^0\Sigma^0$,
$\eta\Lambda$, $\eta\Sigma^0$, $\pi^+\Sigma^-$, $\pi^-\Sigma^+$,
$K^+\Xi^-$, $K^0\Xi^0$. In the following, the meson-baryon channels are
numbered according to this ordering.

\begin{figure}[tbp]
    \centering
    \includegraphics[width=8cm,clip]{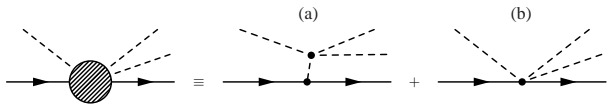}
    \caption{\label{fig:2}
    Basic terms entering the threshold production of pions in $\pi
    N\to \pi\pi N$. (a) : meson pole term, (b) contact term.}
\end{figure}

Concerning the initial process, 
the hatched blob in Fig.~\ref{fig:1}
is given by the sum of meson pole terms and contact terms
as shown in Fig.~\ref{fig:2}.
From the Lagrangian of Eq.~\eqref{eq:MLag} we obtain the meson-meson
amplitude of order $f^{-2}$
and from the Lagrangian of Eq.~\eqref{eq:BLag}
($D$ and $F$ terms), at order $f^{-1}$, the meson-baryon-baryon
coupling as shown in Eq.~\eqref{eq:Yukawa}.
In this way we get an amplitude at lowest order of 
$\mathcal{O}(f^{-3})$, which requires for consistency to calculate the
order $\mathcal{O}(f^{-3})$ from the $D$ and $F$ terms of
Eq.~\eqref{eq:BLag} by expanding $u_{\mu}$ in the number of meson
fields up to three as shown in Eq.~\eqref{eq:Contact}.
This generates the contact term of Fig.~\ref{fig:2} (b).


\subsection{On-shell factorization}

One interesting observation in Ref.~\cite{Oset:1998it} is that the $V$
amplitudes can be factorized on-shell (as a function of $s$) inside
the meson-baryon loops appearing in Fig.~\ref{fig:1}.
In the present case there is also an on-shell factorization for the 
initial process $\pi^-p\to K^0 MB$ as we discuss briefly.  
This process is then followed by the 
final state interaction, shown by the open blob.

\begin{figure}[tbp]
    \centering
    \includegraphics[width=8cm,clip]{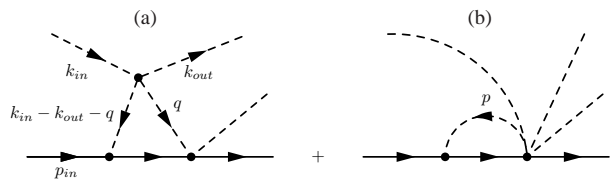}
    \caption{\label{fig:3}
    One loop functions for $MB \to MMB$.}
\end{figure}

For that purpose, let us consider the one loop
diagram of Fig.~\ref{fig:3} (a).
In the following, we first show the factorization of
the meson-meson amplitude with the
momentum $q$ on-shell, then show the cancellation of the off-shell 
part of the meson-meson amplitude associated to
 the momentum $k_{in}-k_{out}-q$.
With these arguments, the on-shell factorization of the meson-baryon
loops~\cite{Oset:1998it} can be applied to the present initial
process, and we can calculate the whole amplitude by evaluating
the initial process at the tree level, separated from the subsequent
meson-baryon loops.

Let us start with showing that the meson-meson amplitude
factorizes 
in Fig.~\ref{fig:3} (a) with the momentum $q$ on-shell.
The $s$-wave meson-meson amplitudes
from the chiral Lagrangians at lowest order
have the form~\cite{Oller:1997ti}
\begin{equation}
    t_{MM}=as+\sum_{i}b_im_i^2+\sum_{i}\beta_{i}(q_i^2-m_i^2)
    \label{eq:MMamp} \ ,
\end{equation}
where the term with $\beta$ gives the off-shell extrapolation of the 
amplitude.
If we take just this off-shell part for the meson of momentum $q$ in 
Fig.~\ref{fig:3} (a), the loop function reads as
\begin{equation}
\begin{split}
    &\int \frac{d^4q}{(2\pi)^4}(q^2-m^2)\frac{1}{q^2-m^2}
    D_{\pi}(k_{in}-k_{out}-q) \\
    &\quad \times G_N(p_{in}+k_{in}-k_{out}-q)\bm{\sigma}
    \cdot(\bm{k}_{in}-\bm{k}_{out}-\bm{q})
\end{split}
    \label{eq:mloop}
\end{equation}
using the non relativistic form $(\bm{\sigma}\cdot \bm{p})$
for the $MBB$ vertex, which will be improved later on to account for 
relativistic corrections.
One should note that in Eq.~\eqref{eq:mloop}, the off-shell part of
the meson-meson amplitude cancels the meson propagator and leads to 
a contracted diagram of the type of Fig.~\ref{fig:3} (b).
On the other hand, it is interesting to note that genuine diagrams of
the type of Fig.~\ref{fig:3} (b) appear from the consideration
of the $BBMMMM$
terms that come from an expansion in the meson fields of the chiral
Lagrangian.
These terms should be added for consistency.
However, by changing $\bm{q}$ to the $\bm{p}$ variable of
Fig.~\ref{fig:3} (b) in the loop functions and realizing that the
dominant term in the $\gamma^{\mu}\partial_{\mu}$ structure of the
$BBMMMM$ vertex comes from the $\gamma^0\partial_{0}$ component (and 
hence no three momentum dependence), the loop functions of
Fig.~\ref{fig:3} vanish at this order (corrections coming at order
$\mathcal{O}(1/2M)$).

There is also a cancellation 
of the off-shell part of the meson-meson amplitude
for the meson with momentum $k_{in}-k_{out}-q$ in Fig.~\ref{fig:3} (a).
It appears already at tree level, but it comes
from an exact cancellation between the off-shell part of the meson
pole term and the contact term.
This fact 
justifies the attempt to find out the on-shell $\pi\pi$
scattering amplitude from analysis of the $\pi N\to \pi\pi N$ data
omitting the contact term~\cite{Lange:2000xu}, except for the contribution 
of other terms in the process~\cite{Oset:1985wt}.
This off-shell cancellation found here is important conceptually.
In practice, we just calculate the meson pole
term with the $p_{ex}$ variable off-shell and add the contact term in
each case, and the cancellation takes place automatically.

Next we look into a possible contribution from the $p$-wave part
of the meson-meson amplitude.
As anticipated, we are looking at the $\pi^-p\to K^0\pi\Sigma$ reaction 
close to threshold of the $\bar{K}N$ production in $\pi^-p\to
K^0\bar{K}N$.
This means that three momenta of the three
final particles in the $\pi^-p\to K^0 MB$ are negligible with respect
to their energies.
Therefore, the on-shell factorization will just pick up the
$s$-wave part of the $MM$ amplitude.
One might argue that the $p$-wave part of the meson-meson amplitude
will not be small when taken inside loops.
By looking again to the diagram of Fig.~\ref{fig:3} (a), the $p$-wave
part of the amplitude would lead to a contribution in the loops of
the type
\begin{equation}
\begin{split}
    &\int d^4q \bm{k}_{in}\cdot \bm{q} D_{\pi}(k_{in}-k_{out}-q)
    D(q) \\
    &\quad \times G(p_{in}-k_{in}-k_{out}-q) \bm{\sigma} \cdot
    (\bm{k}_{in}-\bm{k}_{out}-\bm{q})
\end{split}
    \label{eq:loop} \ .
\end{equation}
Since we know that $k_{in}\sim 1500 \text{ MeV}/c$ and the $q$
integral has a cut off of $600 \text{ MeV}/c$~\cite{Oset:1998it},
$(q/k_{in})^2$ is a small quantity which would allow one to take a
constant propagator for the meson of momentum $k_{in}-k_{out}-q$.
Since $\bm{k}_{out}\sim 0$, the term with
$(\bm{k}_{in}\cdot\bm{q})\,(\bm{\sigma}\cdot\bm{k}_{in})$ vanishes in
the integral, but there remains an integral
\begin{equation}
    \int d^4q \,(\bm{k}_{in}\cdot \bm{q}) \,(\bm{\sigma}
    \cdot\bm{q})
    D(q) G_N(p_{in}-k_{in}-k_{out}-q) 
    \label{eq:loop2} \ ,
\end{equation}
which should be reasonably smaller than the corresponding 
term from the meson meson s-wave which is proportional to 
$\bm{\sigma}\cdot \bm{k}_{in}$.
Yet, there is more to it.
With $\bm{k}_{in}+\bm{p}_{in}=\bm{0}$, and $\bm{k}_{out}\sim 0$, the
argument of $G$ depends on $\bm{q}^2$ and we are left with an
integral of the type
\begin{equation}
    \begin{split}
	&\int d^4q q_iq_j 
	D(q) G_N(p_{in}-k_{in}-k_{out}-q) \\ 
	\sim & \frac{1}{3}\delta_{ij}\int d^4q \bm{q}^2
	D(q) G_N(p_{in}-k_{in}-k_{out}-q)
    \end{split}
    \label{eq:loop3} \ .
\end{equation}
After performing the $q^0$ integration, we are left with an integral
\begin{equation}
    \bm{\sigma} \cdot\bm{k}_{in}
    \int d^3q
    \frac{1}{2\omega(q)}\frac{\bm{q}^2}{M_I-\omega(q)-E(q)+i\epsilon}
    \label{eq:loop4} \ ,
\end{equation}
with $M_I$ the invariant mass of the $MB$ system and $\omega$, $E$
the meson, baryon energies.
The zero in the denominator of Eq.~\eqref{eq:loop4} gets the on-shell
condition for a momentum $q_{on}$ and we can write
\begin{equation}
    \begin{split}
	\bm{q}^2&=\omega(q)^2-\omega(q_{on})^2+\bm{q}_{on}^2  \\
	M_I-\omega(q)-E(q)&=
	\omega(q_{on})-\omega(q)+E(q_{on})-E(q)  
    \end{split}
    \label{eq:momentum}
\end{equation}
By neglecting $E(q^{on})-E(q)$ which holds in the heavy baryon limit
(we are all around neglecting $1/M$ terms), the off-shell part of
Eq.~\eqref{eq:loop4} leads to
\begin{equation}
    \int d^3q \frac{1}{2\omega(q)}\frac{\omega(q)^2-\omega(q_{on})^2}
    {\omega(q_{on})-\omega(q)}
    \sim \int d^3q \frac{\omega(q)^2-\omega(q_{on})^2}
    {\omega(q_{on})^2-\omega(q)^2}
    \label{eq:loop5} \ ,
\end{equation}
which is constant in energy.
This energy independent term, multiplying the
$\bm{\sigma}\cdot\bm{k}_{in}$ factor, can be reabsorbed into, for
instance, the contact term with the use of renormalized coupling
constants, say the physical values of $f_{\pi}$.

\subsection{Factorized amplitude}

With the arguments given above, our approach will require the
evaluation of the meson pole and contact terms for the ten coupled
channels $\pi^-p\to K^0M_iB_i$, using the $s$-wave $MM\to MM$
amplitude where the $\pi^-K^0M_i$ are factorized on-shell.
Since we also saw that the intermediate propagator with momenta
$k_{in}-k_{out}-q$ could also be factorized,
this means we factorize the whole $\pi^-p\to K^0M_iB_i$ amplitude
on-shell outside the loop integral.
The remaining loop function contains only the meson of momentum $q$
and baryon
propagators and this is the $G_{MB}(M_I)$ function found out in the
study of the meson-baryon interaction in Ref.~\cite{Oset:1998it}.
Hence the whole amplitude for the process $\pi^-p\to K^0\pi\Sigma$
corresponding to the upper diagrams of Fig.~\ref{fig:1} is given by
\begin{equation}
    \begin{split}   
	-it_{\text{chiral}}
	=&\bm{\sigma}\cdot\bm{k}_{in}\Bigl[
	(a_{\pi\Sigma}+b_{\pi\Sigma}) \\
	&+\sum_{i}(a_i+b_i)G_i(M_I)t_{i\to \pi\Sigma}(M_I)\Bigr]
    \end{split}
    \label{eq:formula1} \ ,
\end{equation}
where $i$ runs for the ten coupled channels, $t_{i\to\pi\Sigma}$
is the transition T-matrix from the channel $i$ to $\pi\Sigma$
studied in Ref.~\cite{Oset:1998it} and $a_i$, $b_i$ are the on-shell 
contributions to the $\pi^-p\to K^0M_iB_i$ tree level amplitude from
the meson pole and contact terms, respectively.
In the appendix, we give the $a_i$ and $b_i$ terms for all the ten
channels, where, as mentioned before, we have already assumed the
three momenta of the final particle negligible in all channels.

For completeness, we also include a recoil factor from the
$\gamma^{\mu}\gamma_5\partial_{\mu}$ BBM vertex
\begin{equation}
    F_i=\left(1-\frac{p_{ex}^{0(i)}}{2M_p}\right)
    \label{eq:recoil} \ .
\end{equation}
In addition, we also consider the strong form factor of the $MMB$
vertex for which we take a standard monopole form factor for all vertices
\begin{equation}
    F_f(\bm{p})=\frac{\Lambda^2-m_{\pi}^2}{\Lambda^2+\bm{p}^2}
    \label{eq:FF}
\end{equation}
with $\Lambda\sim 800$ MeV.
We take the form factor static to avoid the fictitious poles of the
covariant $(\Lambda^2-m^2)/(\Lambda^2-p^2)$ form.
But we have checked that using this latter form only changes the
results at the level of less than 5 \%.
Given  the cancellation of the off-shell part
of the meson pole term with the contact term, which makes the sum of 
the two terms independent of a possible unitary transformation in the
fields, the form factor is applied both in the meson pole and the
contact term.
This is analogous to what is done with the pion pole and
Kroll-Ruderman term in $\gamma N\to \pi N$ to preserve gauge
invariance~\cite{Carrasco:1992vq}.


\section{Results with the chiral amplitudes}

We perform the calculations for an initial pion momentum
of 1.69 GeV, at which the experiment is done.
The $\pi\Sigma$ invariant mass distribution is given by
\begin{equation}
\begin{split}
    \frac{d\sigma}{dM_I}
    =&\frac{1}{(2\pi)^3}\frac{1}{4s}
    \frac{M\tilde{M}}{\lambda^{1/2}(s,M^2,m_{\pi}^2)}\frac{1}{M_I} \\
    &\times 
    \lambda^{1/2}(s,M_I^2,m_{K}^2)\lambda^{1/2}(M_I^2,\tilde{M}^2,m^2)
    \bar{\Sigma}\Sigma|t|^2
\end{split}
    \label{eq:Mdist} \ ,
\end{equation}
where $M$ and $\tilde{M}$ are the masses of the nucleon 
and the baryon of the final state, 
in this case a $\Sigma$, and $m$ the mass of the final
meson, in this case a $\pi$.

The distributions are calculated for $\pi^+\Sigma^-$, $\pi^-\Sigma^+$
and $\pi^0\Sigma^0$ in the final states. According to the findings of
Ref.~\cite{Nacher:1998mi} in the $\Lambda(1405)$ photoproduction,
the $I=1$ contribution approximately 
cancels in the sum of the $\pi^+\Sigma^-$,
$\pi^-\Sigma^+$ contributions and the $\pi^0\Sigma^0$ does not have
$I=1$ contribution, such that
$\sigma(\pi^+\Sigma^-)+\sigma(\pi^-\Sigma^+)\sim 2
\sigma(\pi^0\Sigma^0)$ and both distributions give the $I=0$
contribution to the process, hence the $\Lambda(1405)$ contribution.

\begin{figure}[tbp]
    \centering
    \includegraphics[width=8cm,clip]{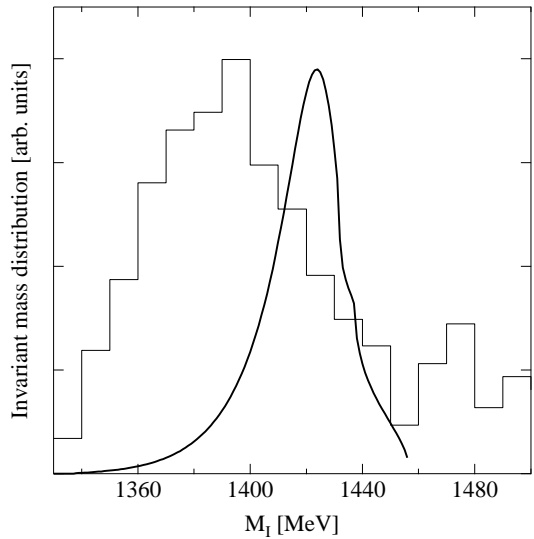}
    \caption{\label{fig:5}
    Contribution of the meson pole and contact terms (diagrams of
    Fig.~\ref{fig:1}) to the $\pi\Sigma$ invariant mass obtained
    averaging $\pi^+\Sigma^-$ and $\pi^-\Sigma^+$.
    The histogram show the
    experimental data taken from Ref.~\cite{Thomas:1973uh}.}
\end{figure}

In Fig.~\ref{fig:5}, we can see the results obtained with these
mechanisms compared to the experimental distribution from
Ref.~\cite{Thomas:1973uh}.
We can see that the theoretical distribution peaks around 1420 MeV
while the experimental one has the peak around 1400 MeV.
The theoretical distribution is also much narrower than experiment.
The disagreement between theory and experiment is apparent.

We can easily trace back the origin of the shape of the theoretical
distribution.
Indeed, the tree amplitude $\pi^-p\to K^0M_iB_i$ for the case of 
$M_iB_i=\bar{K}N$ involve the combinations
$3F-D$ and $D+F$,
which are large compared to the $D-F$ combination that we find for
$M_iB_i\equiv\pi\Sigma$ (we take $F=0.51$ and $D=0.75$).
Therefore, that the sum of the terms in Eq.~\eqref{eq:formula1}
is dominated by the $\bar{K}N$ terms,
giving a larger weight to the
$t_{\bar{K}N\to\pi\Sigma}$ amplitude than to the $t_{\pi\Sigma\to\pi\Sigma}$
one.
As we mentioned, the $\bar{K}N$ states couple strongly to the
$\Lambda(1405)$ resonance of higher energy and weakly to the
$\Lambda(1405)$ of lower energy.
As a consequence, what we see is a distribution which mostly peaks
around the resonance found in Ref.~\cite{Jido:2003cb}
at the pole position $z_R\sim (1426+i16)$ MeV, with a width of
around 30 MeV.
The slightly smaller energy of the peak in Fig.~\ref{fig:5} and
larger width reflects the small contribution of the resonance of
lower energies, also present in the $t_{\bar{K}N\to\pi\Sigma}$, as
well as from the $t_{\pi\Sigma\to\pi\Sigma}$ amplitudes in the sum of 
Eq.~\eqref{eq:formula1}, which are dominated by the lower energy
$\Lambda(1405)$ resonance.
This latter one appears at $z_R=1390+i66$ MeV.

\section{The s-channel resonance contribution}

Since we have $\sqrt{s}\sim 2$ GeV, one could think of the
possibility of having resonance excitation in the $\pi N$ collision
leading to the decay of the resonance in $MMB$.
We would like to have some resonance that can couple to the $MMB$
strongly in $s$-wave. All $S=0$ baryon resonances in the region of
$1700\leq \sqrt{s} \leq 2100$ MeV correspond 
to higher partial waves in the $\pi N$
collision, except for the $N^*(1710)$ and the $N^*(2100)$, which are
$P_{11}$ resonances with the same quantum numbers of the
nucleon~\cite{Hagiwara:2002fs}.
Out of these two, the $N^*(1710)$ resonance has
a very large branching ratio to $\pi\pi
N$ (40-90\%), while the one of the $N^*(2100)$ is unknown, probably 
small, since the large branching ratio seems to be for $N\eta$
(with large errors).
We thus rely upon the $N^*(1710)$ resonance to provide some
contribution to the $\pi^-p\to K^0\pi\Sigma$ process.
Although one can derive different couplings of this resonance to the
$MMB$ in an SU(3) scheme (see Ref.~\cite{Oset:2000ev} for analogy in
other $P_{11}$ resonances), the absence of the kinematically allowed $\eta
\pi N$ channel in the decay mode of the $N^*(1710)$ strongly suggest a
Weinberg-Tomozawa like coupling where this mode is strictly forbidden
at the tree level (see $C_{ij}$ coefficients in
Refs.~\cite{Inoue:2001ip,Hyodo:2003dt}).
This also has the implicit assumption that the $N^*(1710)$
resonance belongs to an SU(3) octet representation, which is the
option adopted in the particle data table~\cite{Hagiwara:2002fs}.
We then assume a coupling of the type of
\begin{equation}
    \mathcal{L}_{N^*\to MMB}
    =\frac{\tilde{B}}{f^2} < \bar{B} i \gamma^{\mu}
[(\Phi \overleftrightarrow{\partial_{\mu}}\Phi) B^*
- B^* (\Phi \overleftrightarrow{\partial_{\mu}}\Phi)] >
    \label{eq:NMMBLag} \ ,
\end{equation}
where now $n^*(1710)$ and $p^*(1710)$ would substitute
in the $B^*$ matrix as the $n$ and $p$. This Lagrangian
is the same that appears in the $s$-wave scattering of
meson-baryon~\cite{Oset:1998it} as we have seen in
Eq.~\eqref{eq:WTint}.
The Lagrangian of Eq.~\eqref{eq:NMMBLag} leads to the amplitude
\begin{equation}
    t_{N^*\to MMB}
    =-\frac{\tilde{B}}{f^2}C_i(\omega_1-\omega_2)
    \label{eq:NMMBint} \ ,
\end{equation}
where $\omega_1$, $\omega_2$ are the energies of the two mesons and
$C_i$ are the $C_{ij}$ SU(3) coefficients found in
\cite{Oset:1998it,Inoue:2001ip,Hyodo:2003dt} and reproduced below in
Table~\ref{tbl:1} for the $N^*$ with zero charge going to pions and
\begin{table}[tbp]
    \centering
    \caption{The $C_i$ coefficients entering Eq.~\eqref{eq:NMMBint}.}
    \begin{ruledtabular}
    \begin{tabular}{cccc}
         & $\pi^-\pi^0 p$ & $\pi^+\pi^- n$ & $\pi^0\pi^0 n$  \\
        \hline
        $C_i$ & $-\sqrt{2}$ & $1$ & $0$  \\
    \end{tabular}
\end{ruledtabular}
    \label{tbl:1}
\end{table}
in Table~\ref{tbl:2} for $N^*$ going to $K^0MB$.
\begin{table}[tbp]
    \centering
    \caption{The $C_i$ coefficients entering Eq.~\eqref{eq:NMMBint}
    with a $K^0$ in the final state.}
    \begin{ruledtabular}
   \begin{tabular}{cccccc}
         $\phantom{\dfrac{1}{1}}$ 
	 & $K^-K^0p$ & $\bar{K}^0K^0 n$ & $\pi^0K^0\Lambda$
	 & $\pi^0K^0\Sigma^0$ & $\eta K^0\Lambda$  \\
        \hline
        $C_i$ & $1$ & $2$ & $-\frac{\sqrt{3}}{2}$ & $\frac{1}{2}$
	& $\frac{3}{2}$ \\
        \hline
	$\phantom{\dfrac{1}{1}}$  & $\eta K^0\Sigma^0$
	 & $\pi^+K^0\Sigma^-$ & $\pi^- K^0\Sigma^+$
	 & $K^+K^0\Xi^-$ & $K^0K^0\Xi^0$  \\
	\hline
	$C_i$ & $-\frac{\sqrt{3}}{2}$ & $1$ & $0$ & $0$ &
	$0$  \\
    \end{tabular}
\end{ruledtabular}
    \label{tbl:2}
\end{table}
The $\tilde{B}$ coefficient is easily derived from the partial decay width
$N^*\to \pi^+\pi^- n, \pi^-\pi^0 p$, where we have
\begin{equation}
    \Gamma_{\pi\pi N}
    =\frac{M M_R}{16\pi^3 \sqrt{s}}
    \int_{\omega_{min}}^{\omega_{max}}d\omega
    \int_{\omega_{min}}^{\omega_{max}}d\omega^{\prime}
    \Theta(1-a^2)\bar{\Sigma}\Sigma|t|^2
    \label{eq:Gamma1} \ ,
\end{equation}
where $\Theta(x)$ is a step function and 
\begin{equation}
    a=\frac{1}{2kk^{\prime}}
    \Bigl\{
    (M_R-\omega-\omega^{\prime})^2
    -M^2-k^2-k^{\prime 2}\Bigr\}
    \label{eq:adef} \ ,
\end{equation}
with $k$, $k^{\prime}$ the moduli of the two pion momenta and 
\begin{equation}
    \bar{\Sigma}\Sigma|t|^2
    =3\frac{\tilde{B}^2}{f^4}(\omega-\omega^{\prime})^2
    \label{eq:pipiNamp} \ .
\end{equation}
Similarly, the $\pi^-p\to N^{*}$ coupling is given by (including the
isospin factor)
\begin{equation}
    -it=\frac{A}{f}\bm{\sigma}\cdot\bm{k}_{in}
    \label{eq:piNamp} \ ,
\end{equation}
by means of which the partial decay width is given by
\begin{equation}
    \Gamma_{\pi N}
    =\frac{1}{2\pi}\frac{M}{\sqrt{s}}\frac{A^2}{f^2}k_{in}^3
    \label{eq:Gamma2} \ .
\end{equation}
Assuming the middle values of the $N^*$ width 
($\Gamma \sim 100$ MeV)
and partial decay widths for 
$\pi\pi N$ and $\pi N$ channels ($\Gamma_{\pi\pi N}=65$ MeV 
and $\Gamma_{\pi N}=15$ MeV), we find
\begin{equation}
    |A|=0.086 \ , \quad
    \tilde{|B|}=0.77
    \label{eq:ABcoef1} \ .
\end{equation}
For later convenience, we refer to this parameter set 
including $M_R=1710$ MeV as Set I.

\begin{figure*}[tbp]
    \centering
    \includegraphics[width=16cm,clip]{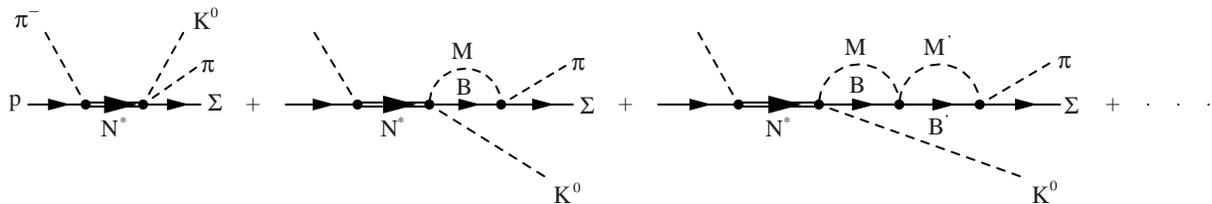}
    \caption{\label{fig:6}
    Resonant mechanisms for $\Lambda(1405)$ production in the
    $\pi^-p\to K^0\pi\Sigma$ reaction.}
\end{figure*}

The $\pi^- p\to K^0\pi\Sigma$ production \textit{via} 
$N(1710)$ and $\Lambda(1405)$
production is now given diagrammatically in Fig.~\ref{fig:6}.
The amplitude for $\pi\Sigma$ production is now given by
\begin{equation}
\begin{split}
    -it_R
    =&\frac{A}{f}\bm{\sigma}\cdot\bm{
    k}_{in}
    \frac{i}{\sqrt{s}-M_R+i\frac{\Gamma}{2}}
    (-i)\frac{(-)\tilde{B}}{f^2} \\
    &\times\Bigl[
    C_{\pi\Sigma}(\omega_{\pi}-\omega_{K^0}) \\
    &+\sum_{i}C_i(\omega_i-\omega_{K^0})G_i(M_I)t_{i\to \pi\Sigma}\Bigr]
\end{split}
    \label{eq:Ramp} \ ,
\end{equation} 
with $\pi\Sigma=\pi^0\Sigma^0, \pi^-\Sigma^+, \pi^+\Sigma^-$,
and $\omega_i$, $\omega_{K^0}$ given by their on-shell values,
following the same arguments used in Ref.~\cite{Oset:1998it}
to factorize the $MB\to MB$ amplitude on-shell in the loops,
\begin{equation}
    \omega_{K^0}=\frac{s+m_K^2-M_I^2}{2\sqrt{s}} \ ,
    \quad 
    \omega_i=\frac{M_I^2+m_i^2-M_i^2}{2M_I} \ ,
    \label{eq:energy}
\end{equation}
with $m_i, M_i$ the meson and baryon masses of the particle in the
$N^*\to K^0MB$ reaction, and $M_I$ the $\pi\Sigma$ invariant mass.
Furthermore, in Eq.~(\ref{eq:Ramp}) $\Gamma$ is the total width 
whose energy dependence is taken into account by using 
Eqs.~(\ref{eq:Gamma1}) and (\ref{eq:Gamma2}) for the 
$\pi \pi N$ and $\pi N$ channels, respectively, and 
by considering a $k^3$ dependence for the $\eta N$ channel.

\section{Results with the resonant mechanism : Final results}

\begin{figure}[tbp]
    \centering
    \includegraphics[width=8cm,clip]{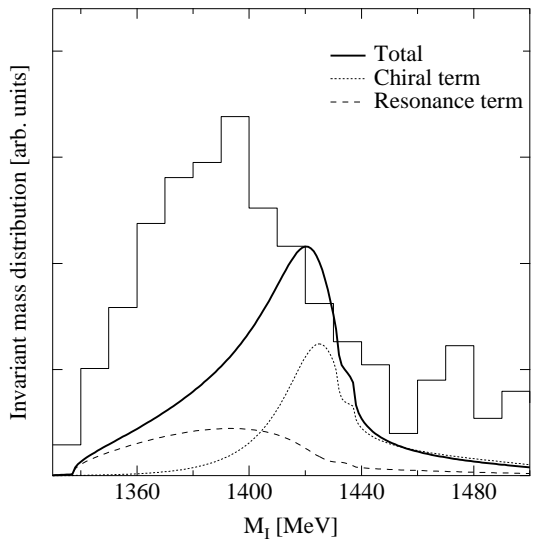}
    \caption{\label{fig:155mdist100}
    Invariant mass distribution of $\pi\Sigma$ obtained by averaging
    $\pi^+\Sigma^-$ and $\pi^-\Sigma^+$ with parameter Set I.
    The histogram shows the
    experimental data taken from Ref.~\cite{Thomas:1973uh}}
\end{figure}

In Fig.~\ref{fig:155mdist100}, we show the results 
that we obtain for the
resonant mechanism (dashed line) with Set I,
together with
the results obtained before from the
chiral mechanisms (dotted line).
The calculation was performed at the energy 
$\sqrt{s} = 2020$ MeV, or equivalently 
$p_\pi = 1690$ MeV in the laboratory frame.
This is 
the energy at which the experiment we compare with 
was done~\cite{Thomas:1973uh}.
Although the figure is shown in arbitrary units, we have adjusted the
relative scale between the experimental and theoretical curves assuming 
that the integrated experimental mass distribution should coincide with 
the total cross sections in the $\pi \Sigma$ channels given in 
Ref.~\cite{Thomas:1973uh}.  
Theoretical and experimental total cross sections for various channels 
are shown in Table~\ref{tbl:sigma}.
We can see that the strength of the resonant mechanism is smaller 
than that of the chiral terms, 
however, the $\pi\Sigma$ distribution created by the resonant
mechanism is much broader and peaks around 1390 MeV.
It is instructive to see the reason for the shape of the resonant
mechanism.
Indeed, we have seen that the $N^*\to NM_1M_2$ vertex goes like
$\tilde{B}(\omega_1-\omega_2)$.
Now for the case of the $K^0\bar{K}N$ channel, the amplitude goes
like $\omega_{K^0}-\omega_{\bar{K}}$, but we are at low energies,
close to the $K^0\bar{K}N$ threshold production, where the difference
of the two kaon energies is close to zero.
On the other hand, in the $N^*\to K^0\pi\Sigma$, the difference
between the $K^0$and $\pi$ energies is finite and of the order of 300
MeV in the region that we study.
Hence, the $K^0 \pi\Sigma$ channel is strongly favored and according 
to Eq.~\eqref{eq:Ramp}, the final $\pi\Sigma$ production channel is
practically given by $t_{\pi\Sigma\to\pi\Sigma}$.
The factors $(\omega_i-\omega_{K^0})$ and $G_i(M_I)$ give extra
weight to this amplitude to finally produce a distribution that
essentially reflects the lower energy $\Lambda(1405)$ resonance to
which the $\pi\Sigma$ channel couples strongly.

The coherent sum of the two mechanisms, taking $A\cdot\tilde{B}> 0$, 
leads to a mass distribution,
given by the solid line, which
still remains to be dominated by the chiral terms and the 
agreement with the data is not very good.  

\begin{figure}[tbp]
    \centering
    \includegraphics[width=8cm,clip]{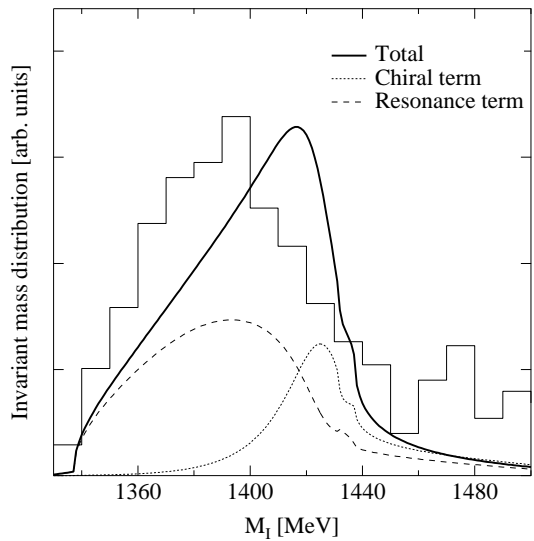}
    \caption{\label{fig:155mdist200}
    Same as in Fig.~\ref{fig:155mdist100} but 
    with Set II.}
\end{figure}

It is possible to improve the theoretical mass distribution 
if we play a bit with uncertainties in the 
resonance mass, the total $N^*$ width and the branching ratios. 
By assuming $M_R \sim 1740$ MeV, 
$\Gamma=200$ MeV and $\Gamma_{\pi N}=40$ MeV	and
$\Gamma_{\pi\pi N}=100$ MeV (We refer to this parameter set as Set II)
well within the experimental boundaries, we obtain the results of
Fig.~\ref{fig:155mdist200} where the agreement with the data 
becomes acceptable.
The increase in the resonant part is mostly 
due to the increase in the 
$\pi NN^*(1710)$ coupling constant when using the larger partial
width $\Gamma_{\pi N}=40$.  
In Table \ref{tbl:sigma}, we have summarized cross sections of various 
channels comparing experimental data and theoretical results with
the two sets of parameters.
Except for the $K^0 \pi^0 \Lambda$ channel in which 
$\Sigma(1380)$ resonance, not accounted for in our study,
plays a major role, the agreement between 
theory and experiment is acceptable for Set II.
We can also see that the use of Set II not only improves the mass
distribution but also the global agreement with the individual
cross sections.
Note the importance of the interference in the chiral 
and resonant terms in order to obtain a better agreement 
between theory and experiment.  

\begin{table}[tbp]
    \centering
    \caption{Total cross sections for several final states 
    with parameter sets I and II in units of [$\mu$b].
    Experimental data are taken from Ref.~\cite{Thomas:1973uh}.}
    \begin{ruledtabular}
    \begin{tabular}{clllll}
	    final state & $K^0K^-p$ & $K^0\bar{K}^0n$ & $K^0\pi^0\Lambda$
	    & $K^0\pi^+\Sigma^-$ & $K^0\pi^-\Sigma^+$  \\
	    \hline
	    chiral  & 2.36 & 2.84 & \phantom{00}3.14 &
	    \phantom{0}3.04 &
	    \phantom{0}6.78  \\
	    resonance(I) & 0.29 & 0.28 & \phantom{00}4.47 & 
	    \phantom{0}6.68 & \phantom{0}2.24  \\
	    resonance(II) & 0.70 & 0.67 & \phantom{0}10.85 & 16.18 &
	    \phantom{0}5.43  \\
	    total(I) & 2.82 & 4.61 & \phantom{00}1.93 &
	    12.00 & 14.31  \\
	    total(II) & 3.75 & 5.98 & \phantom{00}6.02 &
	    21.32 & 20.01  \\ \hline
	    Exp.  & 2.9 & 8.3 & 104.0 & 25.1 & 20.2  \\
	\end{tabular}
	\end{ruledtabular}
    \label{tbl:sigma}
\end{table}


\section{Conclusions}

We have developed a model for the $\pi^-p \to K^0\pi\Sigma$ reaction 
in the region of excitation of the $\Lambda(1405)$ resonance.
We discussed the fact that present theoretical models using chiral
dynamics and coupled channel unitarization are all converging to the
existence of two poles close to the nominal $\Lambda(1405)$ resonance.
They would reflect the singlet and one $I=0$ octet (although with
some mixture), which are dynamically generated in these approaches.
The two resonances appear at different energies and couple very
differently to the $\pi\Sigma$ and $\bar{K}N$ channels.

When we try to construct a model for the $\pi^-p\to K^0\pi\Sigma$
in analogy to the low energy chiral model for $\pi N\to \pi\pi N$,
we observe that the chiral model stresses the role of $\bar{K}N$
intermediate state making
the total amplitude for $\pi^-p\to K^0\pi\Sigma$
mostly sensitive to the $t_{\bar{K}N\to \pi\Sigma}$ amplitude, which
is dominated by a narrow resonance peaking at 1426 MeV.
The mechanism alone leads to a $\pi\Sigma$ mass distribution in strong 
disagreement with the experimental data.

On the other hand, it was found that there are complementary
mechanisms exciting
$N^*$ resonances from the $\pi^-p$ entrance channel.
Inspection of the partial waves involved in the resonance excitations 
and the decay modes singled out a resonance which gives
contribution to the 
process, the $N^*(1710)$, with the same quantum numbers of the
nucleon.
The strong $N\pi\pi$ decay channel together with the absence of the 
$N\pi\eta$ channel, suggested a coupling of the $N^*(1710)$ resonance
to $BMM$ of
the SU(3) Weinberg-Tomozawa type, which we exploited to see the
consequences in
the $\pi^-p\to K^0\pi\Sigma$ reaction.
We observed that this new mechanism had an opposite behavior to the 
chiral one, and strongly stressed the $\pi\Sigma$ intermediate state 
instead of the $\bar{K}N$, leading to a production amplitude
dominated by the $t_{\pi\Sigma\to\pi\Sigma}$ amplitude.
Since this amplitude is dominated by the wide resonance peaking
around 1390 MeV, we found that the $\pi\Sigma$ mass distribution
roughly followed the shape of this resonance and was wide and peaking
at an energy below 1400 MeV. 
The coherent sum
of the two mechanism was shown to lead to total cross sections and 
a mass distribution compatible
with the experiment, within the theoretical and experimental 
uncertainties.

The exercise done here shows the important role played by the two
resonance poles in the production process of the nominal
$\Lambda(1405)$ resonance.
We could see how two different mechanisms (chiral and $N^*(1710)$
terms) filtered each one of the
resonance contributions,
and then how the coherent sum of the amplitudes from the two
mechanisms could describe the data.
The present exercise has shown the non-triviality
of the $\Lambda(1405)$ generation,
which has been given for granted in all
previous theoretical studies.
Indeed, one needs to make a
careful theoretical study of each reaction 
in order to understand the nature of the resonance from
the observed shape of the $\pi\Sigma$ mass distribution.

The study is also telling us that there might be other processes where 
the reaction mechanism
of $\Lambda(1405)$ production  filters
one or another resonance, hence leading to very different shapes for
the $\pi\Sigma$ mass distribution.
The $K^-p\to\Lambda(1405)\gamma$ reaction was advocated as one where 
the narrow higher energy resonance will be populated.
The findings of this paper should stimulate further theoretical and
experimental work that helps us pin down the existence and properties
of these two resonances.

\begin{acknowledgments}
This work is supported by the Japan-Europe (Spain) Research
Cooperation Program of Japan Society for the Promotion of Science
(JSPS) and Spanish Council for Scientific Research (CSIC), which
enabled E.~O. and M.~J.~V.~V. to visit RCNP, Osaka and T.~H. and 
A.~H. to visit
IFIC, Valencia.
This work is also supported in part  by DGICYT
projects BFM2000-1326, BFM2001-01868, FPA2002-03265,
the EU network EURIDICE contract
HPRN-CT-2002-00311,
and the Generalitat de Catalunya project
2001SGR00064.
\end{acknowledgments}


\appendix
\section{Amplitudes of chiral terms}

Here we show the tree level amplitudes of chiral terms $a_i$ and $b_i$
in Eq.~\eqref{eq:formula1}.
Indices $i$ are assigned for
$K^-p$, $\bar{K}^0n$, $\pi^0\Lambda$, $\pi^0\Sigma^0$,
$\eta\Lambda$, $\eta\Sigma^0$, $\pi^+\Sigma^-$, $\pi^-\Sigma^+$,
$K^+\Xi^-$, $K^0\Xi^0$ in that order.
Note that for the meson pole term of channel $1(K^-p)$,
both $\pi^0$ and $\eta$ exchange can happen, so that
we show both of them.
\begin{equation}
\begin{split}
    a_1^{(\pi)}
    =&-\frac{1}{4\sqrt{2}f^3}(D+F)
    \frac{1}{(m_{K^0}+q^0-k^0_{in})^2
    -|\bm{k}_{in}|^2-m_{\pi^0}^2} \\
    &\times \Bigl(
    m_{K^0}(m_{K^0}-2k^0_{in})
    -(q^0)^2 + 2k^0_{in}q^0\Bigr)
\end{split}
    \nonumber
\end{equation}
\begin{equation}
\begin{split}
    a_1^{(\eta)}
    =&-\frac{1}{36\sqrt{2}f^3}(-D+3F)
    \frac{1}{(m_{K^0}+q^0-k^0_{in})^2
    -|\bm{k}_{in}|^2-m_{\eta}^2}  \\
    &\times\Bigl(
    m_{K^0}^2-4m_{\pi^-}^2
    +12m_{K^0}q^0 
    +3(q^0)^2 \\
    &+6m_{K^0}k^0_{in}+6q^0 k^0_{in}\Bigr)
\end{split}
    \nonumber
\end{equation}
\begin{equation}
    b_1=\frac{1}{12\sqrt{2}f^3}
    (D-3F)
    \nonumber
\end{equation}
\begin{equation}
\begin{split}    
    a_2
    =&-\frac{1}{6\sqrt{2}f^3}(D+F)
    \frac{1}{(m_{K^0}+q^0-k^0_{in})^2
    -|\bm{k}_{in}|^2-m_{\pi^+}^2} \\
    &\times\Bigl(
    3m_{K^0}^2-2m_{K^0} k_{in}^0
    +4k_{in}^0 q^0
    +2m_{K^0} q^0
    -(q^0)^2
    \Bigr)
\end{split}
    \nonumber
\end{equation}
\begin{equation}
    b_2=-\frac{1}{12\sqrt{2}f^3}
    (D+F)
    \nonumber
\end{equation}
\begin{equation}
\begin{split}    
    a_3
    =&-\frac{1}{4\sqrt{6}f^3}(D+3F)
    \frac{1}{(m_{K^0}+q^0-k^0_{in})^2
    -|\bm{k}_{in}|^2-m_{K^+}^2} \\ 
    &\times\Bigl(
    m_{\pi^-}^2-(q^0)^2+2m_{K^0} q^0
    +2m_{K^0} k^0_{in}
    \Bigr)
\end{split}
    \nonumber
\end{equation}
\begin{equation}
    b_3=-\frac{\sqrt{6}}{48f^3}
    (D+3F)
    \nonumber
\end{equation}
\begin{equation}
\begin{split}
    a_4
    =&\frac{1}{4\sqrt{2}f^3}(D-F)
    \frac{1}{(m_{K^0}+q^0-k^0_{in})^2
    -|\bm{k}_{in}|^2-m_{K^+}^2} \\
    &\times\Bigl(
    m_{\pi^-}^2-(q^0)^2+2m_{K^0} q^0
    +2m_{K^0} k^0_{in}
    \Bigr)
\end{split}
    \nonumber
\end{equation}
\begin{equation}
    b_4=\frac{1}{8\sqrt{2}f^3}
    (D-F)
    \nonumber
\end{equation}
\begin{equation}
\begin{split}
    a_5
    =&\frac{1}{36\sqrt{2}f^3}(D+3F)
    \frac{1}{(m_{K^0}+q^0-k^0_{in})^2
    -|\bm{k}_{in}|^2-m_{K^+}^2} \\
    &\times\Bigl(
    6m_{K^0} k^0_{in}
    -6m_{K^0}q^0
    -8m_{K^0}^2 \\
    &-12q^0 k^0_{in}
    +5m_{\pi^-}^2
    +3(q^0)^2\Bigr) 
\end{split}
    \nonumber
\end{equation}
\begin{equation}
    b_5=-\frac{1}{24\sqrt{2}f^3}
    (D+3F)
    \nonumber
\end{equation}
\begin{equation}
\begin{split}
    a_6=&-\frac{1}{12\sqrt{6}f^3}(D-F)
	\frac{1}{(m_{K^0}+q^0-k^0_{in})^2
	-|\bm{k}_{in}|^2-m_{K^+}^2} \\
	&\times\Bigl(
	6m_{K^0} k^0_{in}
	-6m_{K^0}q^0
	-8m_{K^0}^2 \\
	&-12q^0 k^0_{in}
	+5m_{\pi^-}^2
	+3(q^0)^2\Bigr) 
\end{split}
    \nonumber
\end{equation}
\begin{equation}
    b_6=\frac{\sqrt{6}}{48f^3}
    (D-F)
    \nonumber
\end{equation}
\begin{equation}
    a_7=0 
    \nonumber
\end{equation}
\begin{equation}
    b_7=0 
    \nonumber
\end{equation}
\begin{equation}
\begin{split}
    a_8
    =&-\frac{1}{6\sqrt{2}f^3}(D-F)
    \frac{1}{(m_{K^0}+q^0-k^0_{in})^2
    -|\bm{k}_{in}|^2-m_{K^0}^2} \\
    &\times\Bigl(
    -2m_{K^0} k_{in}^0
    -2k_{in}^0 q^0
    +3m_{\pi^-}^2
    -4m_{K^0}q^0
    -(q^0)^2\Bigr)
\end{split}
    \nonumber
\end{equation}
\begin{equation}
    b_8=\frac{1}{6\sqrt{2}f^3}
    (D-F)
    \nonumber
\end{equation}
\begin{equation}
    a_9=0 
    \nonumber
\end{equation}
\begin{equation}
    b_9=0 
    \nonumber
\end{equation}
\begin{equation}
    a_{10}=0 
    \nonumber
\end{equation}
\begin{equation}
    b_{10}=0 
    \nonumber
\end{equation}

\end{document}